\documentclass[twocolumn,amsmath,amssymb,floatfix,nofootinbib,superscriptaddress]{revtex4}
\usepackage{epsfig}
\newcommand{\beq}{\begin{equation}}
\newcommand{\eeq}{\end{equation}}

\newcommand{\be}{\begin{equation}}
\newcommand{\ee}{\end{equation}}

\begin{document}

\title{Correlations between Ultrahigh Energy Cosmic Rays and AGNs\\ 
%{\it (short author version)} 
}

\author{Glennys R. Farrar, Ingyin Zaw}
\affiliation{Center for Cosmology and Particle Physics \&
Department of Physics\\ New York University, NY, NY 10003, USA
}
\author{Andreas A. Berlind}  
\affiliation{Department of Physics and Astronomy, Vanderbilt University, Nashville, TN 37235, USA}


\begin{abstract}
We investigate several aspects of the correlations reported by the Pierre Auger Observatory between the highest energy cosmic rays (UHECRs) and galaxies in the Veron-Cetty Veron (VCV) catalog of AGNs.  First, we quantify the extent of the inhomogeneity and impurity of the VCV catalog.  Second, we determine how the correlation between the highest energy Auger UHECRs and VCV galaxies is modified when only optically-identified AGNs are used.  Finally, we measure the correlation between the published Auger UHECRs and the distribution of matter.  Our most important finding is that the correlation between UHECRs and AGNs is too strong to be explained purely by AGNs tracing the large scale distribution of matter, indicating that (barring the correlation being a statistical fluke) some substantial fraction of UHECRs are produced by AGNs.  We also find that once we take into account the heavy oversampling of the VCV catalog in the Virgo region, the lack of UHECR events from that region is not incompatible with UHECR having AGN sources.  

\end{abstract}
\maketitle

\section{Introduction}
The Pierre Auger Observatory has reported\cite{augerScience07,augerLongAGN} a correlation between the arrival directions of the highest energy cosmic rays, and the positions of galaxies in the Veron-Cetty Veron ``Catalog of  Quasars and Active Galactic Nuclei" (12th Ed.)\cite{VCV} (VCV).  Using UHECR data above 40 EeV through May 27, 2006 and  VCV galaxies out to $z=0.024$, a prescription was established by scanning on UHECR energy threshold, maximum angular separation, and maximum VCV object redshift.  Then, independent data taken after May 27, 2006 was used to test this prescription with the same energy threshold, maximum redshift and angular separation. When satisfied in mid-summer 2007, the prescription implied an {\it a priori} probability of less than 1\% that UHECR sources are distributed isotropically.  Altogether, in the full dataset up to Aug. 31, 2007,  there are 27 UHECRs with energy above 57 EeV and 20 of them are within 3.2$^{\circ}$ of VCV galaxies with $z < 0.018$ (about 75 Mpc). Restricting to $|b| \geq 10^\circ$, where the VCV catalog is more complete, there are 22 UHECRs of which 19 are correlated.

The existence of a correlation between UHECRs and VCV galaxies clarifies the nature of ultrahigh energy cosmic rays by establishing that UHECRs are extragalactic in origin and that they have an energy-dependent horizon consistent with the GZK prediction.  However as stressed in refs.\cite{augerScience07,augerLongAGN}, the observed correlation alone is insufficient to conclude that the sources of UHECRs are AGNs, because AGNs are clustered with galaxies and thus may only be a tracer of the true sources.  We show here that the observed correlation is too strong to be explained merely by AGNs being clustered along with other galaxies.  

The VCV catalog is not complete, homogeneous, or pure.   We begin by quantifying these deficiencies,  in section \ref{VCV}, in order to understand the limitations of VCV for establishing quantitative bounds on the contribution of AGNs to the population of UHECRs.  Next, using the Auger scan method, we measure in section \ref{scans} the correlation between the published Auger UHECRs above 57 EeV and the optically-identifiable AGNs in VCV, and compare that to the correlation with the full VCV catalog.  Finally, in section \ref{2MRS} we answer the question of whether the observed degree of correlation to VCV and its optically-identified AGNs can be explained just on the basis of the fact that galaxies are clustered and AGNs trace that clustering: we find that the correlation of VCV with published  Auger UHECRs above 57 EeV has less than 1\% chance of arising simply from VCV tracing the distribution of all galaxies.

%the correlation with active galaxies has only a 1\% {\em a priori} probability for sources in galaxies.

\section{ VCV purity and completeness}
\label{VCV}
The VCV catalog is a compilation of the AGN, BL Lac, and quasar candidates reported in the literature, with heterogeneous selection criteria and different amounts of telescope time devoted to different targets.  It is not complete,  uniform or pure. Unfortunately, the required complete, uniform, pure and well-characterized AGN catalog does not exist at this time, so in order to understand the significance of correlations with this catalog, we need to quantify its incompleteness, impurity and non-uniformity of the VCV catalog.  

Zaw, Farrar and Greene\cite{zfg08} (ZFG below) addressed the purity issue, with respect to optical AGN indicators, by examining all (21) of the VCV galaxies with $z\leq 0.018$ which fall within 3.2$^\circ$ of an Auger UHECR above 57 EeV; this examination is extended here as reported below.   ZFG determined that 14 of the 21 VCV galaxies which correlate with UHECRs are optically-identifiable AGNs.  The remaining 7 show limited or no nuclear activity based on their optical spectra.  However only about half of the AGNs found using X-ray or radio can be identified using the standard BPT optical criteria\cite{reviglioHelfand06}, so some of the 7 correlated VCV galaxies which are not optical AGNs could have active nuclei nevertheless.  Observations are in progress with the Chandra X-ray observatory to determine whether these galaxies show signs of AGN activity.

We have extended the work of ZFG by examining the spectra of the VCV galaxies with $z \leq 0.024$, out to $6^\circ$ from each UHECR, to determine which are optically-identifiable AGNs.  There are 66 such VCV galaxies.  Of these, 52 are AGNs by optical criteria\cite{zfg08}.  Nineteen of 20 galaxies labeled S1 or S1.x (broad-line Seyfert) by VCV meet the AGN criteria.  This is not surprising since S1 identifications are much more secure than S2 (narrow-line Seyfert) and other identifications\cite{zfg08}.  On the other hand, none of the 4 galaxies labeled as H2 among the 66 VCV galaxies examined meet optical AGN criteria.  In sum, after galaxies identified as H2 regions are removed, about 1/6 of the galaxies in VCV are not optically-identifiable AGNs; however some of these may have active nuclei which would have to be discovered in the radio or X-ray.

Having estimated the uncertainty in the purity of VCV, we next measure the uniformity of its coverage.  To do this we compare the surface density of the VCV catalog (with galaxies labeled as H2 removed), to that of a complete, volume-limited catalog of galaxies.  2MASS\cite{2MASS} imaged the whole sky in the near-infrared ($J$, $H$, and $K$ bands) and the 2MASS extended source catalog contains positions and fluxes of all galaxies on the sky down to a limiting $K$-band magnitude of 13.5\cite{Jarrett2MASS}.  This waveband provides an excellent, unbiased measure of the distribution of matter in the Universe except where it is incomplete due to Galactic extinction.  Recently, Huchra et al. have compiled redshifts for the $K<11.25$ brightest of these galaxies. We use this 2MASS Redshift Survey (2MRS) catalog to compare with VCV.  

\begin{figure*}[p]
\begin{center}%\noindent $\bullet$ {\bf
\noindent
\epsfig{file=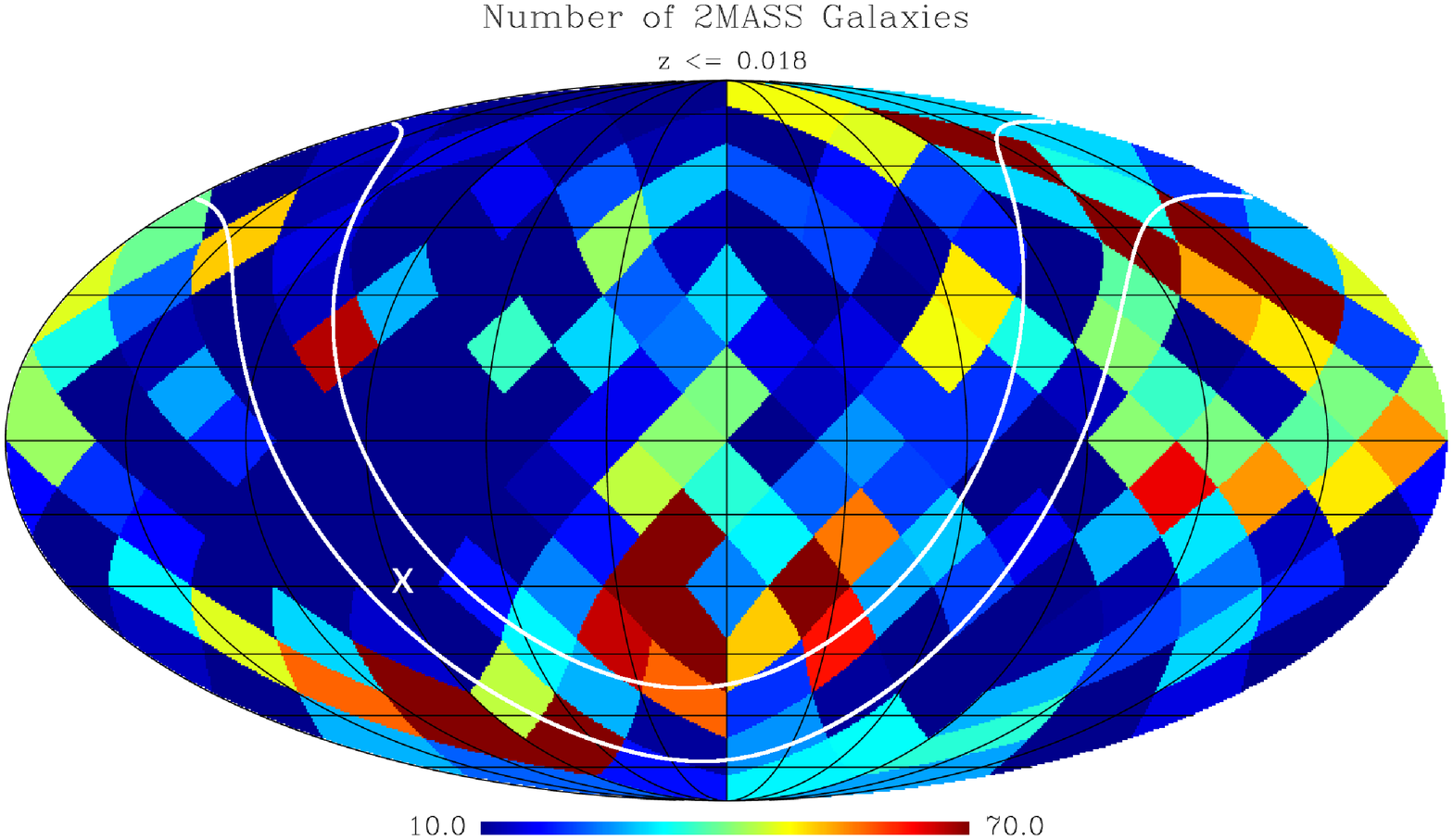,width=3.5 in}
\epsfig{file=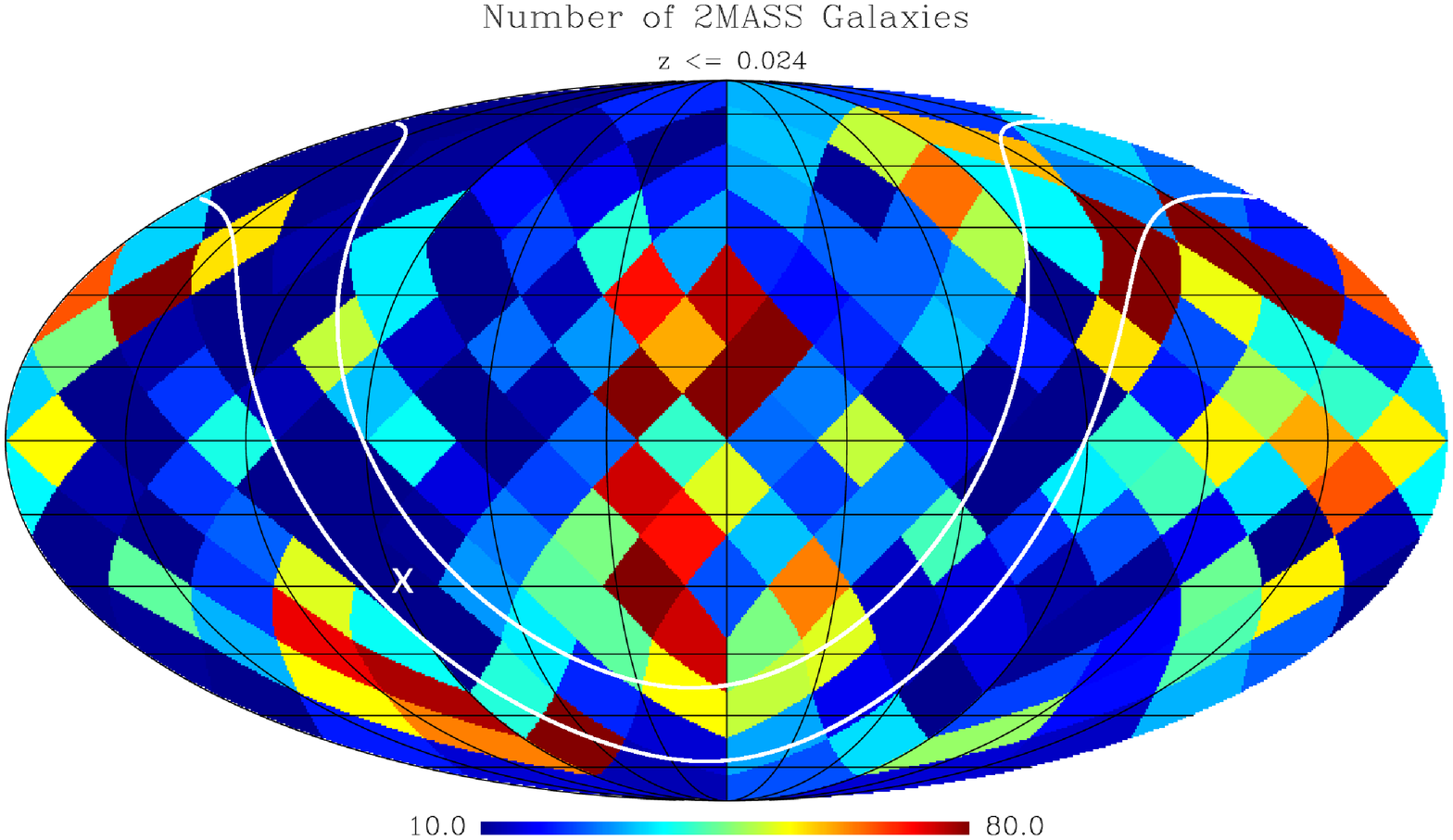,width=3.5 in}
\epsfig{file=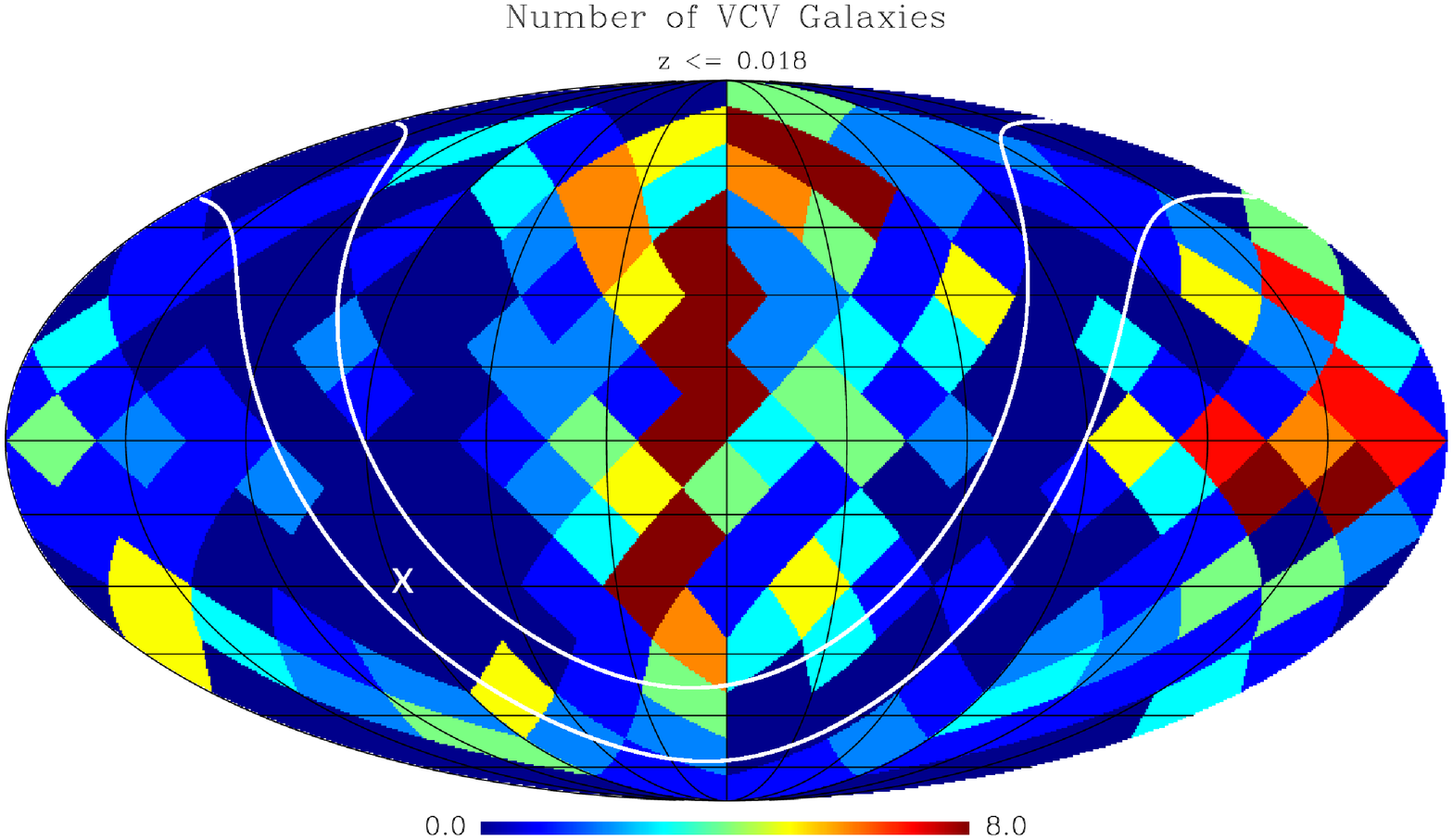,width=3.5 in}
\epsfig{file=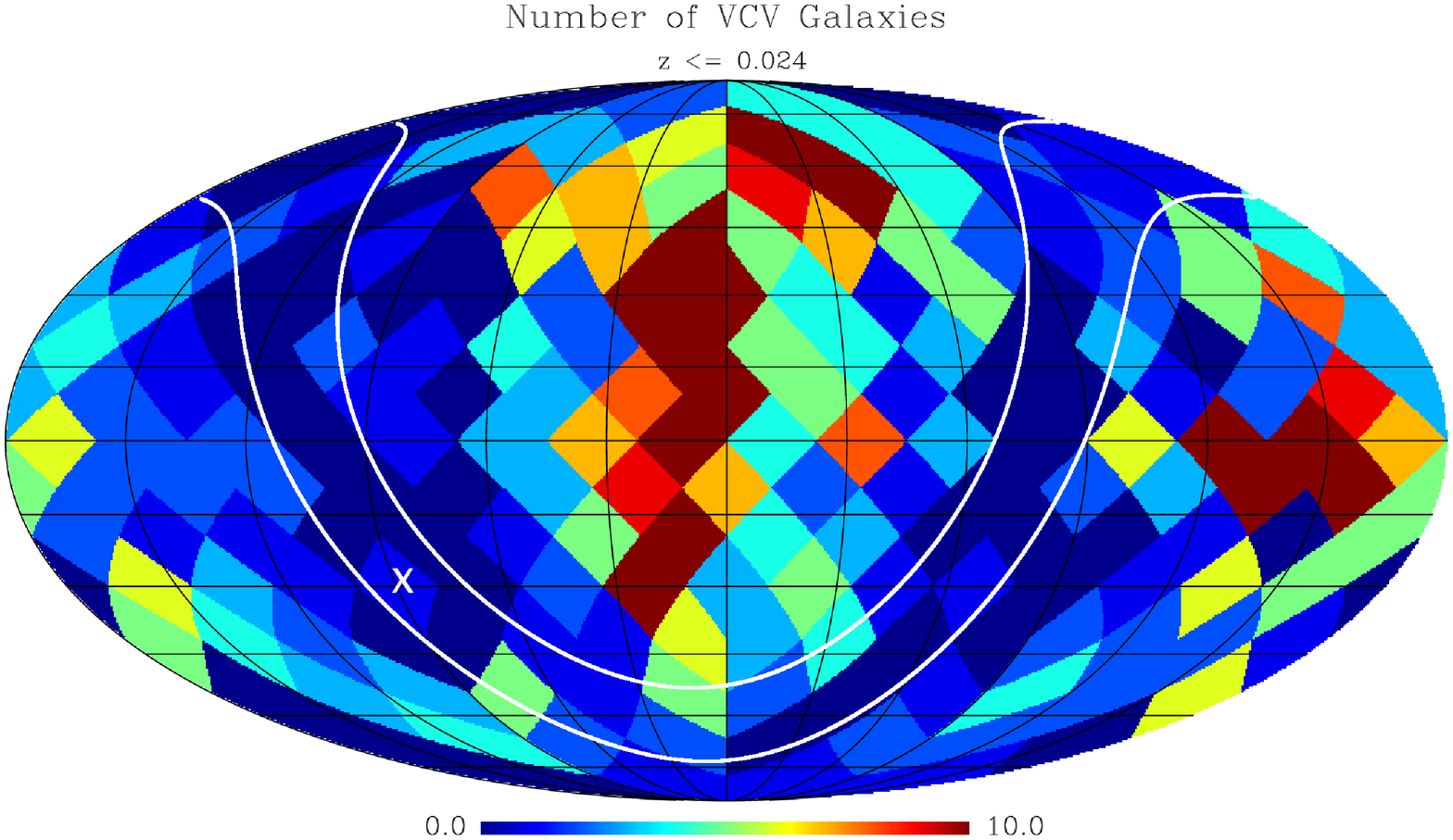,width=3.5 in}
\epsfig{file=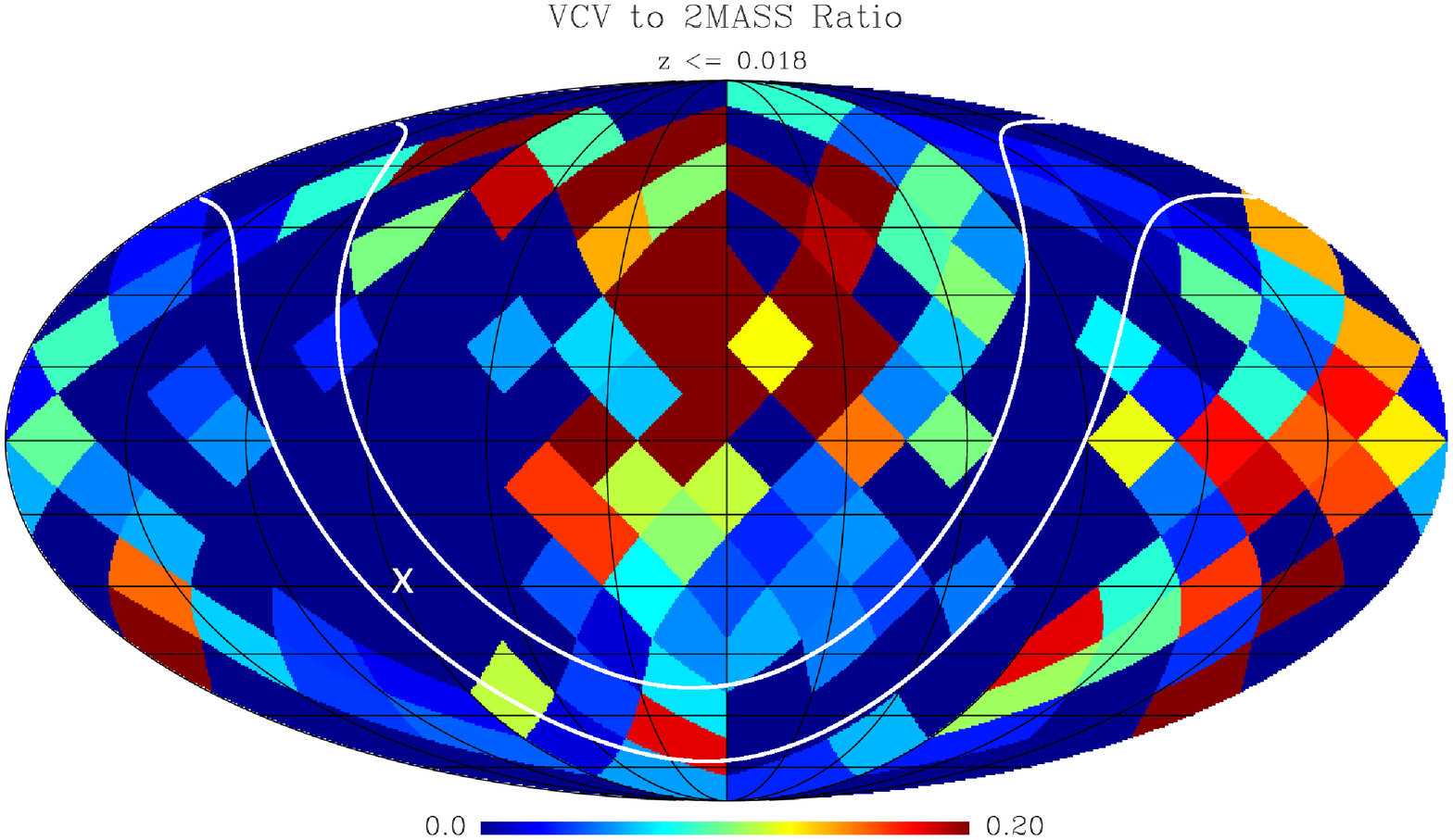,width=3.5 in}
\epsfig{file=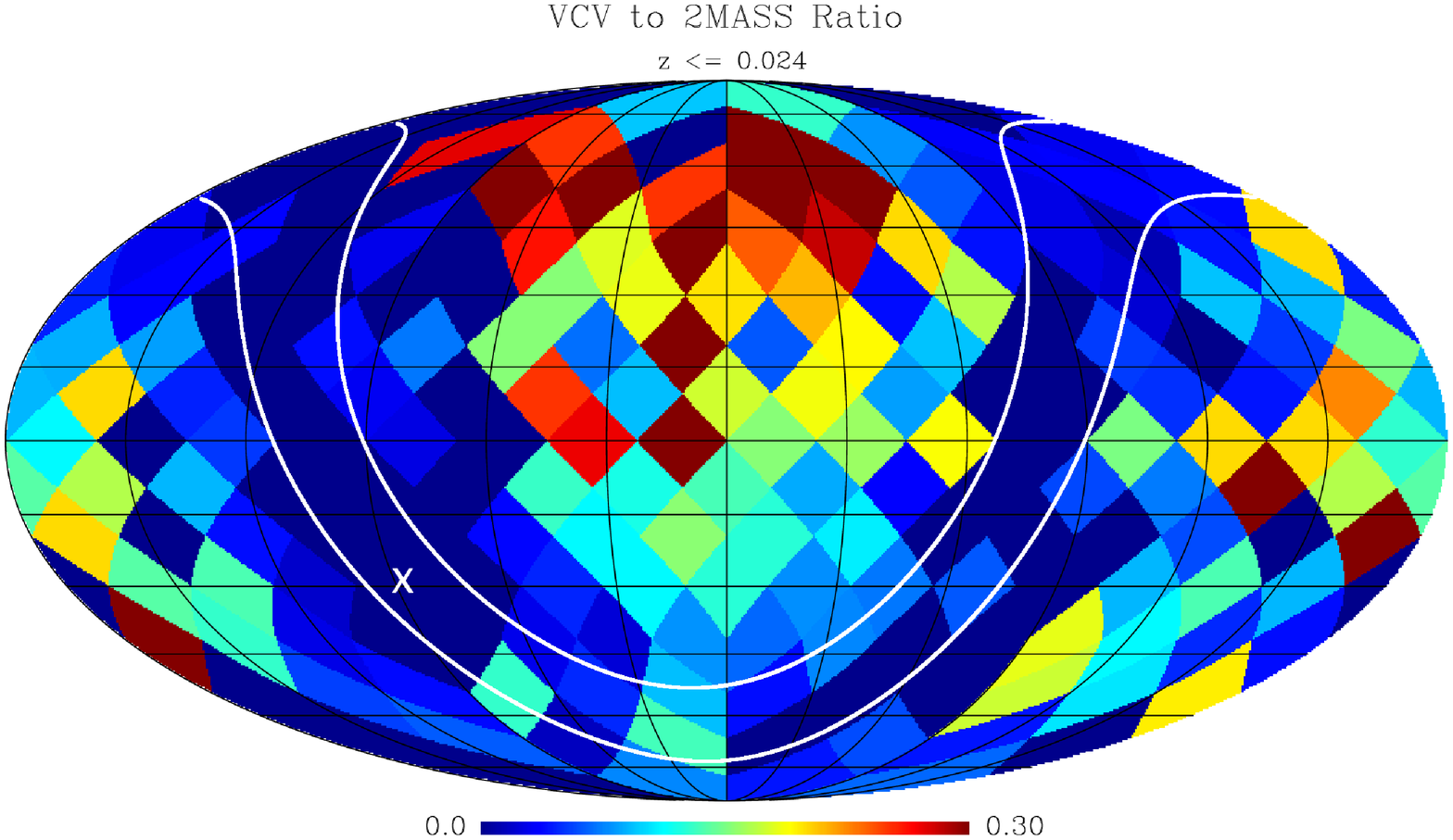,width=3.5 in}
\end{center}
\caption{HEALPix maps of the surface density distribution of 2MRS galaxies (top) and VCV galaxies (middle).  The white lines are at $b = \pm 10^\circ$ and the cross is at the Galactic center.  The bottom panels show the ratio of VCV to 2MRS galaxies, for $z \leq 0.018$ (left) and $z \leq 0.024$ (right), showing the large variation in completeness of VCV.  The maps are in Equatorial Coordinates with RA=180$^\circ$ at the center of the map, using the Mollwide projection. Each map has 192 equal area pixels. The Virgo cluster is centered on RA=187$^\circ$, Dec = 13$^\circ$. }
\label{HealpixMaps}
\end{figure*} 

We divide the sky into 192 equal area patches of $\approx 200$ sq-degrees each, using HEALPix\cite{healpix} (currently, http://healpix.jpl.nasa.gov) and find the number of VCV and 2MRS galaxies in each patch.  The upper and middle panels of Fig. \ref{HealpixMaps} show the surface densities of 2MRS galaxies and of VCV galaxies in equatorial coordinates, for $z \leq 0.018$ on the left and $z \leq 0.024$ on the right.  There are 417 (624) VCV and 5097 (6214) 2MRS galaxies in the respective samples with $z\leq 0.018 \, (\leq 0.024)$, for average VCV/2MRS ratios 0.08 (0.10).  The paucity of both 2MRS and VCV galaxies near the Galactic center in particular and in the Galactic plane in general is due to dust extinction in the Galaxy.  The lower panels show the ratio of the number of VCV to 2MRS, or dark blue (zero) if there are no VCV galaxies in the pixel.  The uncertainty in the ratio due to Poisson noise can be estimated from the number count maps provided in the upper panels.  

It is apparent that relatively more AGN-like galaxies have been identified in the northern hemisphere than in the southern hemisphere, which is not surprising in view of the Sloan Digital Sky Survey's sky coverage and the dedicated search by Ho et al \cite{Ho1997b} for very low-luminosity AGN in the northern hemisphere.   In addition to this large scale difference between hemispheres, the plots in Fig. \ref{HealpixMaps} show that the pixel-to-pixel variation in AGN detection is very large.  The VCV/2MRS ratio ranges from between 0\% and 77\% for $z \leq 0.018$, and between 0\% and 91\% for $z \leq 0.024$.  The VCV/2MRS ratio is $\leq 1$\% in 32\% (22\%) of the pixels, with $\sim$90\% of the patches having ratios below 30\% for $z<0.018$ and 20\% in the deeper sample.   This large degree of inhomogeneity in the AGN sampling of VCV means that even if AGNs (or another type of galaxy that is concentrated in VCV) are the sources of UHECRs, the distribution of UHECR arrival directions cannot be expected to mirror the distribution of VCV galaxies.  In particular, VCV is highly over-sampled in the Virgo region, by a factor of $\sim 5$ within a $10^\circ$ radius of the center of the Virgo Cluster for $z \leq 0.018$ and $\sim 3$ for $z \leq 0.024$, so the absence of UHECR events from this region is compatible with the UHECR-VCV correlation seen in other regions, within statistical fluctuations.  

\begin{figure}[h]
\begin{center}
\noindent
\epsfig{file=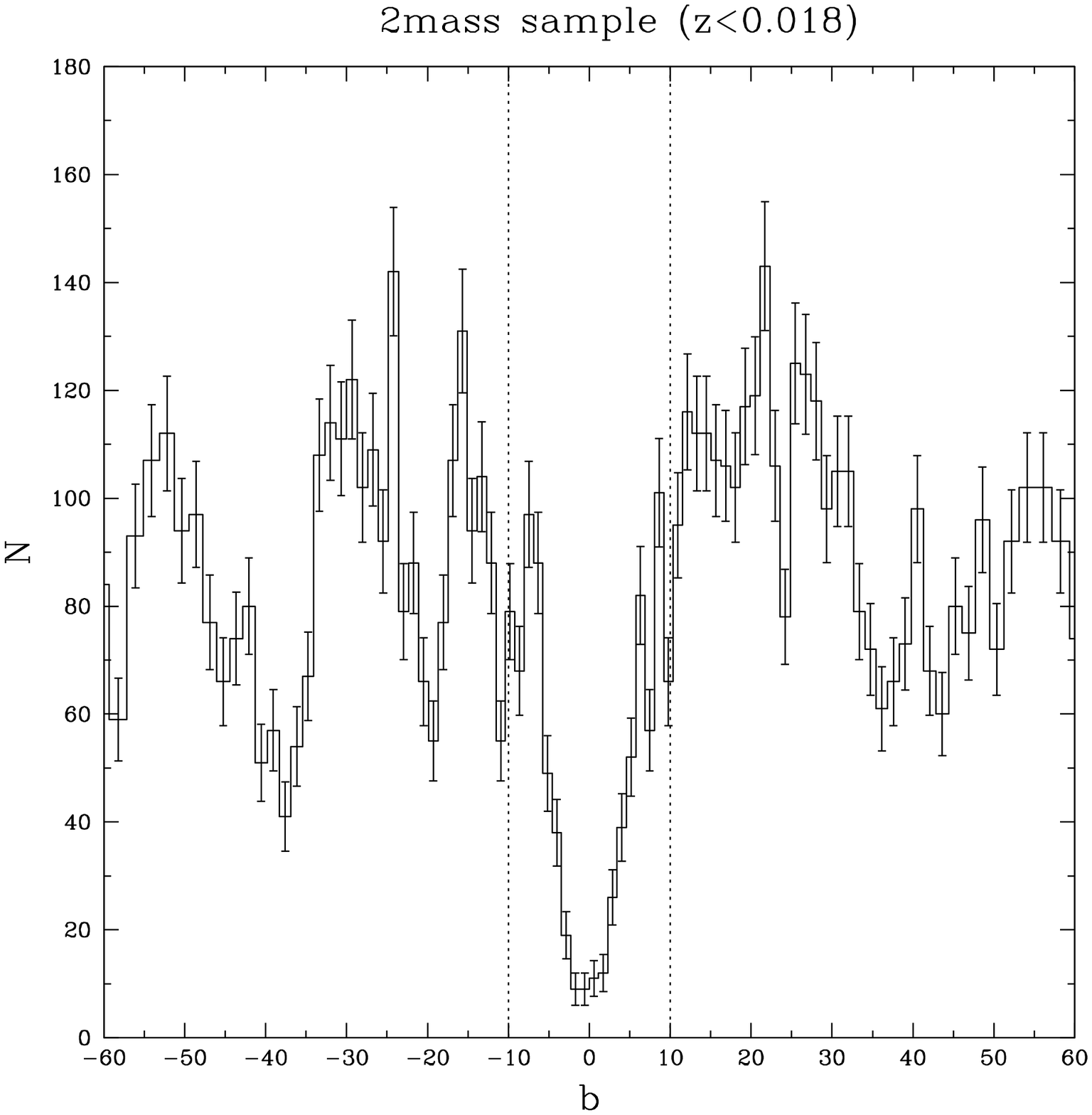,width=3 in}
\end{center}
\caption{
Distribution of 2MRS galaxies within $z<0.018$, as a function of Galactic latitude.}
\label{galcut}
\end{figure}

For correlation studies, it is important to limit the analysis to regions in which the galaxy catalog is complete.  As shown in Fig. \ref{galcut}, which plots the 2MRS galaxies with $z\leq 0.018$ versus $b$, Galactic extinction is an important effect for $|b| \lesssim 10^\circ$.  At larger latitudes, volume-selected subsamples of this catalog should provide a good map of the matter in the nearby universe.  Therefore, we restrict our analyses below to $|b|\geq 10^{\circ}$ and $z\leq0.024$, where the 2MRS K=11.25 catalog contains 11851 galaxies.  We will work with two volume-limited 2MRS catalogs, with $z \leq 0.018$ and $z \leq 0.024$; these catalogs have $M_K \leq -23.2$ and $M_K \leq -23.8$, respectively.  Altogether, there are 405 (606) VCV galaxies and 4525 (5516) 2MRS galaxies in these volume-limited sub-catalogs within these redshift limits and with $|b|>10^\circ$.  Thus on average, there are about 0.09 (0.11) VCV galaxies for each 2MRS galaxy in the volume limited samples, with $|b|>10^\circ$ and in the redshift ranges of interest.  

\begin{figure*}[p]
\begin{center}%\noindent $\bullet$ {\bf
\noindent
\epsfig{file= 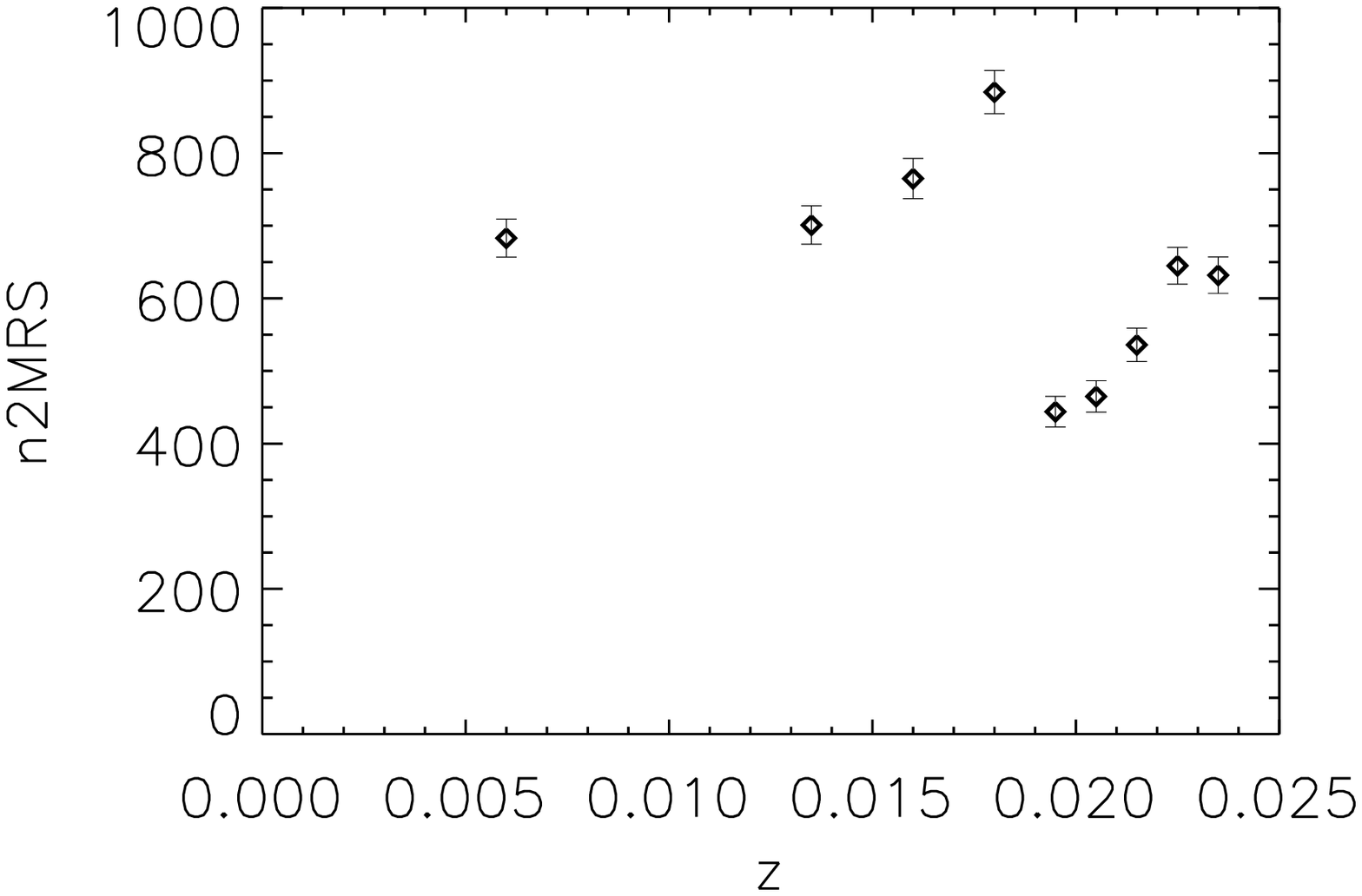,width=2 in}
\epsfig{file= 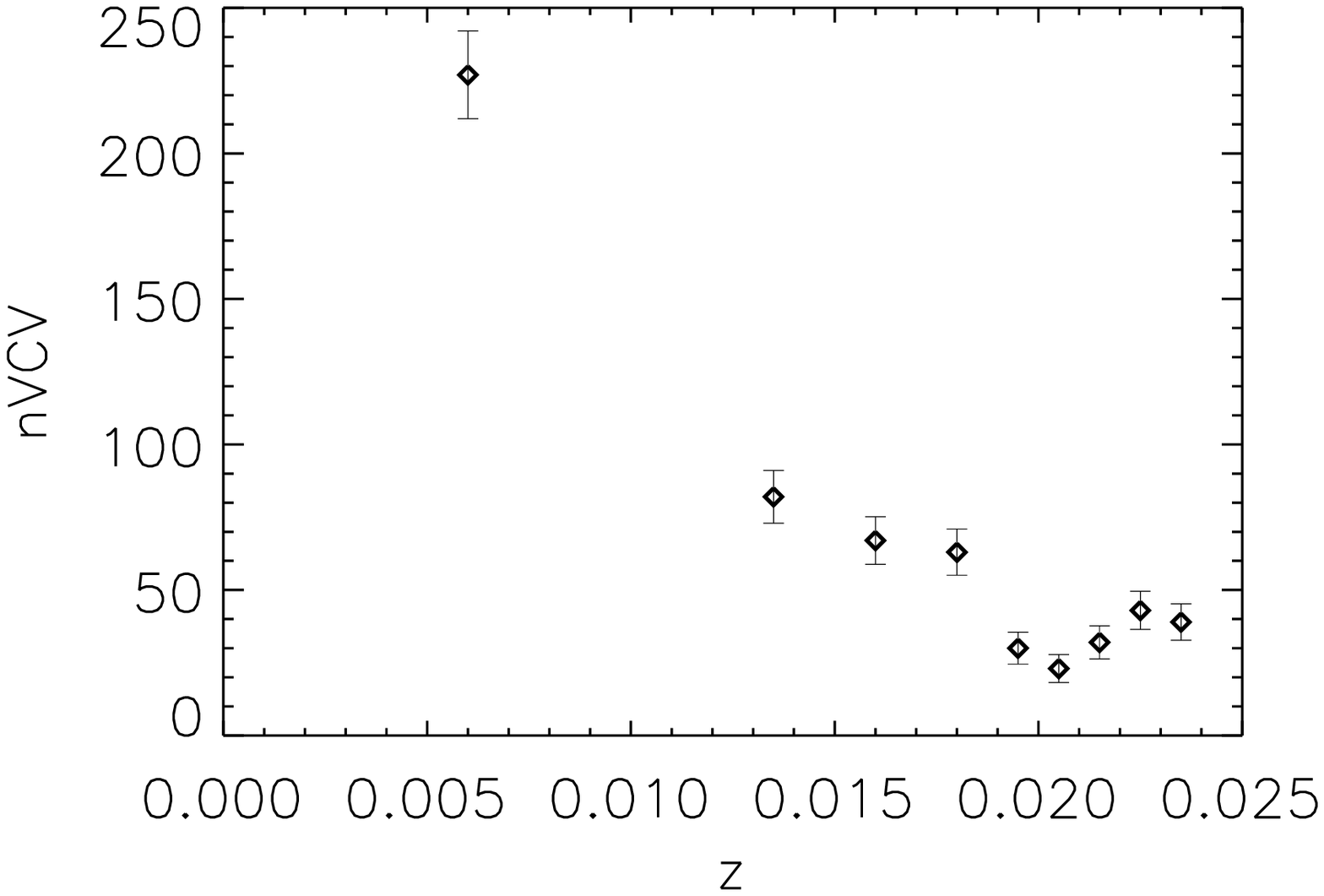,width=2 in}
\epsfig{file=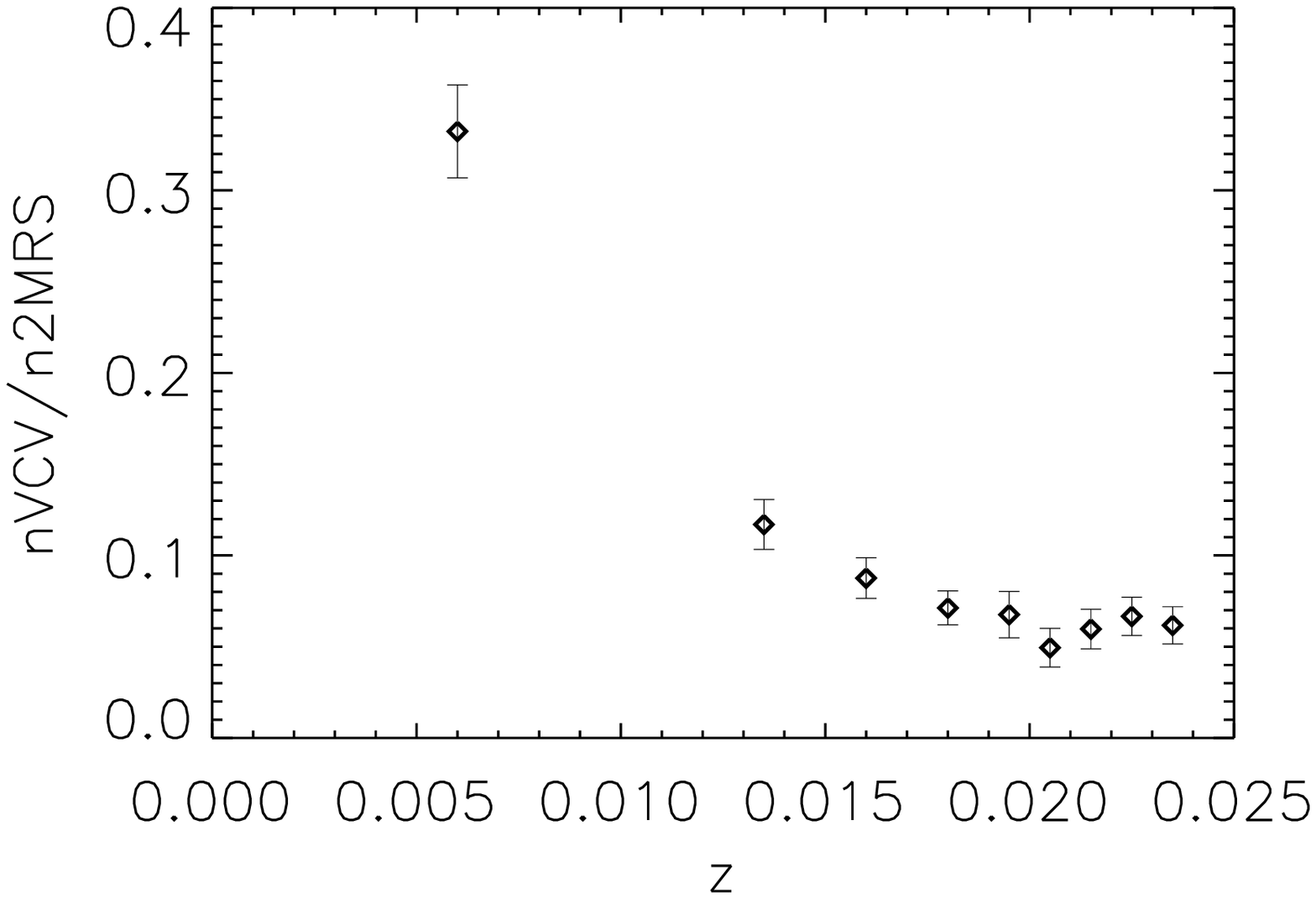,width=2 in}
\epsfig{file= 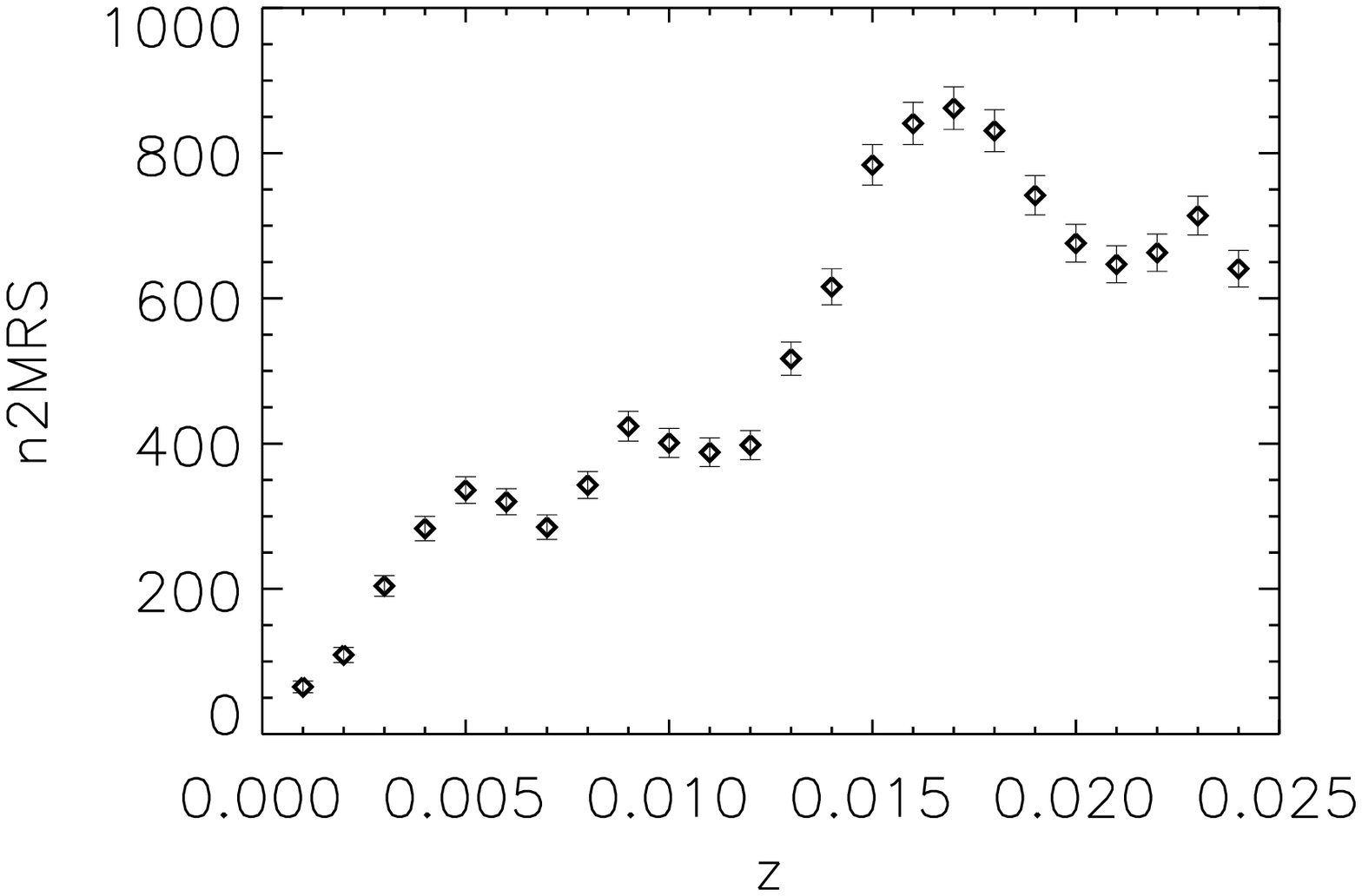,width=2 in}
\epsfig{file= 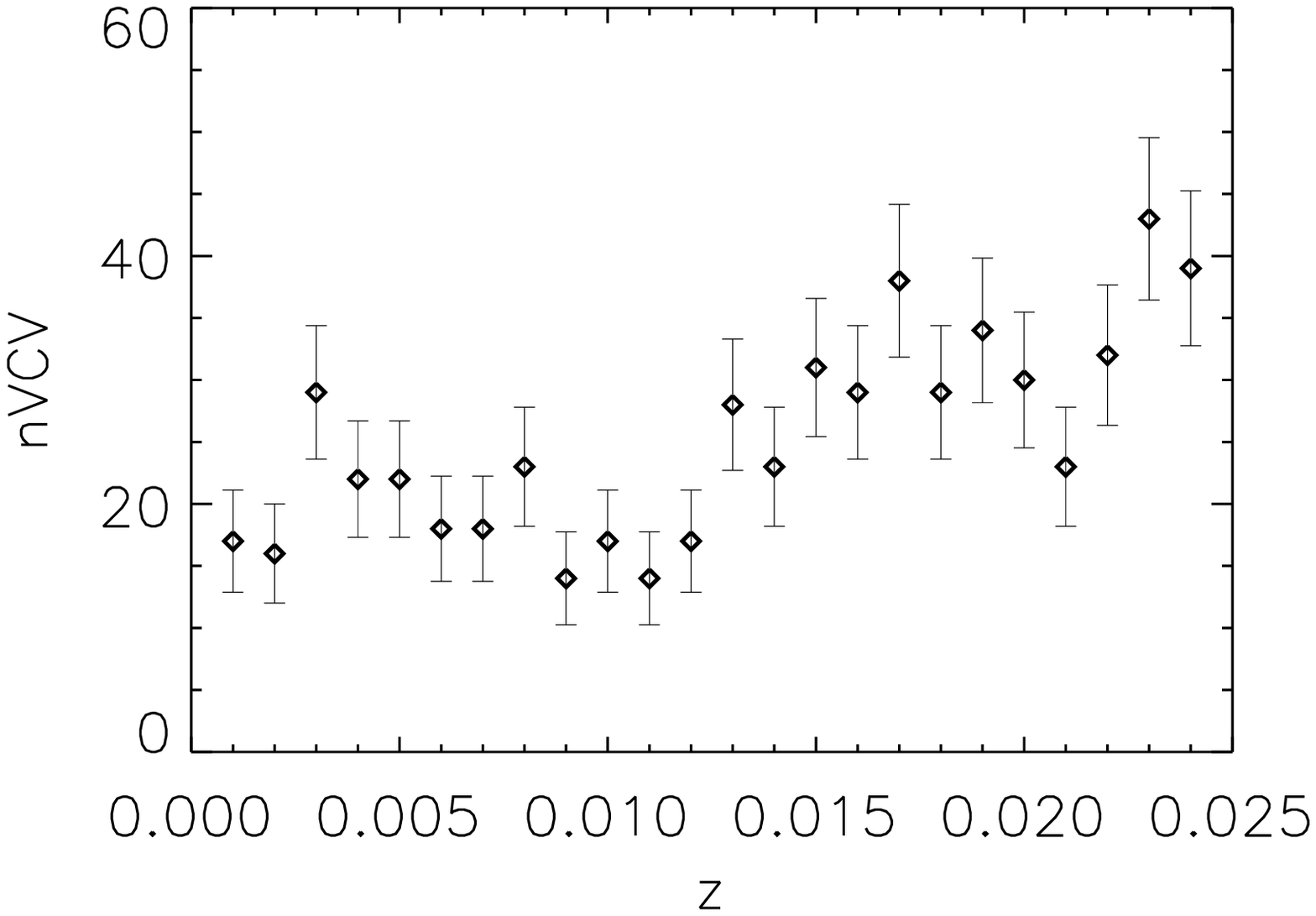,width=2 in}
\epsfig{file=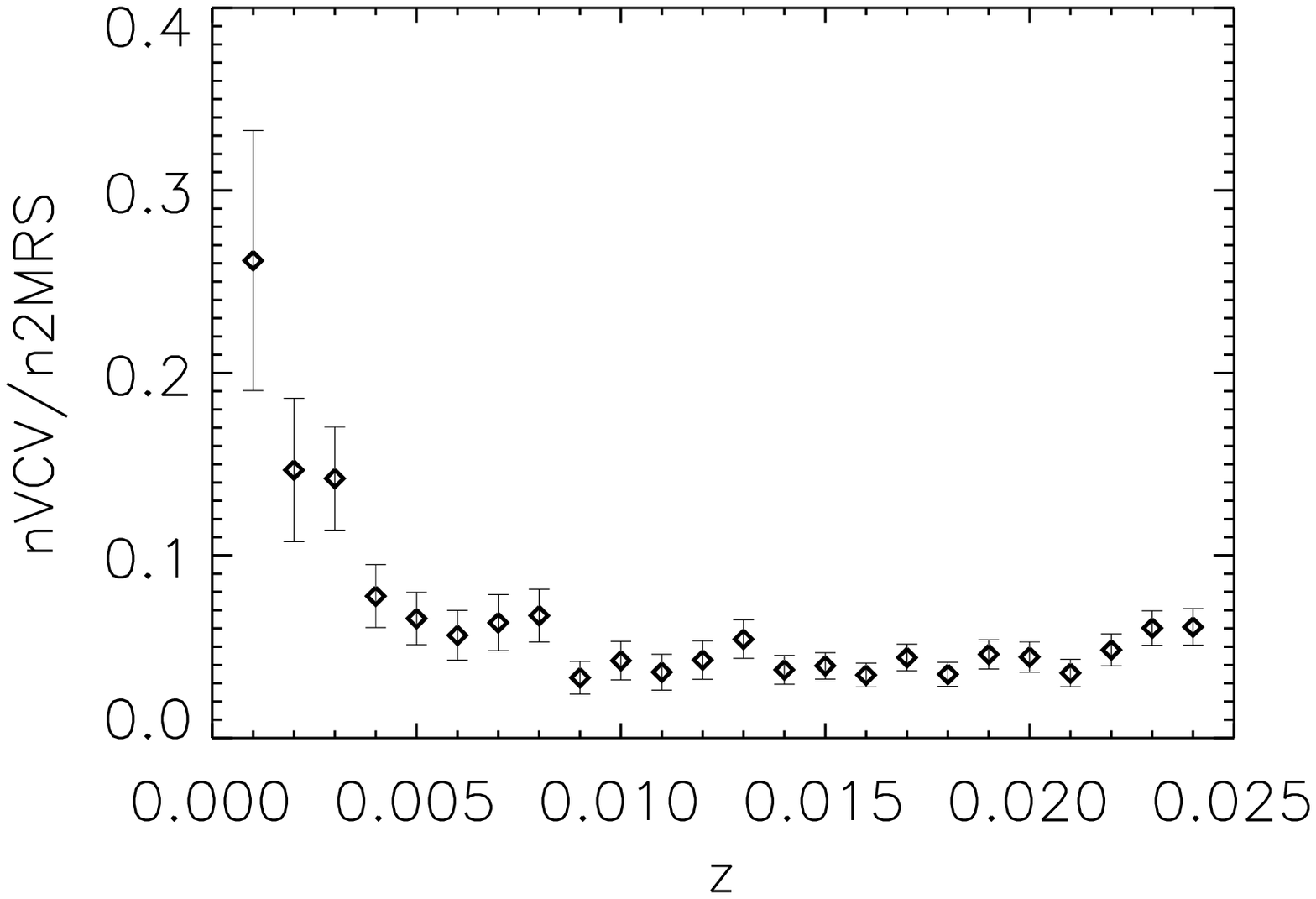,width=2 in}
\end{center}
\caption{Distribution of 2MRS galaxies with $|b| \geq 10^\circ$, in a volume-limited (top left) and a flux-limited (bottom left) sample to $z \leq 0.024$.  Equal-volume and equal-$\Delta z$ bins are used respectively, to have reasonable numbers of events in each bin.  The center column contains the same plots for VCV galaxies and the right column plots show the ratio of the number of VCV galaxies to the number of volume-limited (above) and flux-limited (below) 2MRS galaxies.}
\label{VCVz}
\end{figure*}

Fig. \ref{VCVz}, compares the density distribution of VCV galaxies (with those labeled H2 by VCV removed and  $|b| \geq 10^\circ$), to volume-limited and flux-limited samples of 2MRS galaxies out to $z \leq 0.024$.  The volume-limited sample is binned in equal volume and the flux-limited sample is binned in equal $\Delta z$, in order to have meaningful statistics in each bin. One sees that VCV is over-sampled at low $z$ in comparison to both the flux-limited or volume-limited 2MRS catalogs.

\section{ UHECR correlations with VCV}
\label{scans}
We set the stage for our studies by reviewing the Auger scan analysis using the 27 published UHECR events and scanning over the maximum angular separation, $\psi$, and over the maximum redshift of the galaxies, imposing a Galactic latitude cut $|b| \geq 10^\circ$.   We scan on source distance and maximum angular separation, $\psi$, but not on energy threshold, because lower energy UHECR events are not publicly available.  There are 694 galaxies in the VCV catalog with $z<0.024$ of which 674 have $|b|>10^\circ$, and our UHECR dataset consists of the 22 published Auger events above 57 EeV\cite{augerLongAGN} with $|b|>10^\circ$.  We scan in steps of $0.1^\circ$ and $\Delta z = 0.001$. For each combination of $z_{max}$, and $\psi$, we determine $p$, the exposure-weighted projection of disks of radius $\psi$ around each AGN in the sample, normalized to the total Auger exposure.  From the number of correlated UHECRs, $k_{\rm corr}$, we compute the associated probability measure $P$, which following Auger is the cumulative binomial probability to find $k_{\rm corr}$ correlated events out of $N_{\rm CR}$ cosmic rays, given the exposure-weighted fractional coverage $p$ of AGNs.  The minimum value of $P$ in the scan analysis with VCV galaxies occurs for $z_{max}=0.018$, and $\psi=3.2^\circ$.  For these parameter values, $19$ UHECRs correlate with AGNs and the exposure-weighted fraction of sky covered by VCV galaxies is $p=0.2343$.  This results in a value of the probability measure of $P=7.7 \times 10^{-10}$.  

Following Auger\cite{augerScience07,augerLongAGN}, the chance probability of the correlation is assessed by making many mock catalogs of UHECRs with 22 arrival directions chosen at random according to the Auger exposure, and then performing the same scan on $\psi$ and $z_{\rm max}$.  The resultant probability of finding as low or lower value of $P$ by chance, for an isotropic source distribution, is $(7.1 \pm 2.3) \times 10^{-8}$.  (Note that the values of various quantities quoted above are not identical to the corresponding quantities in the Auger analysis because of our restriction on $|b| \geq 10^\circ$ and the absence of the scan on energy.)
 
\begin{figure}[t]
\begin{center}
\noindent
\epsfig{file=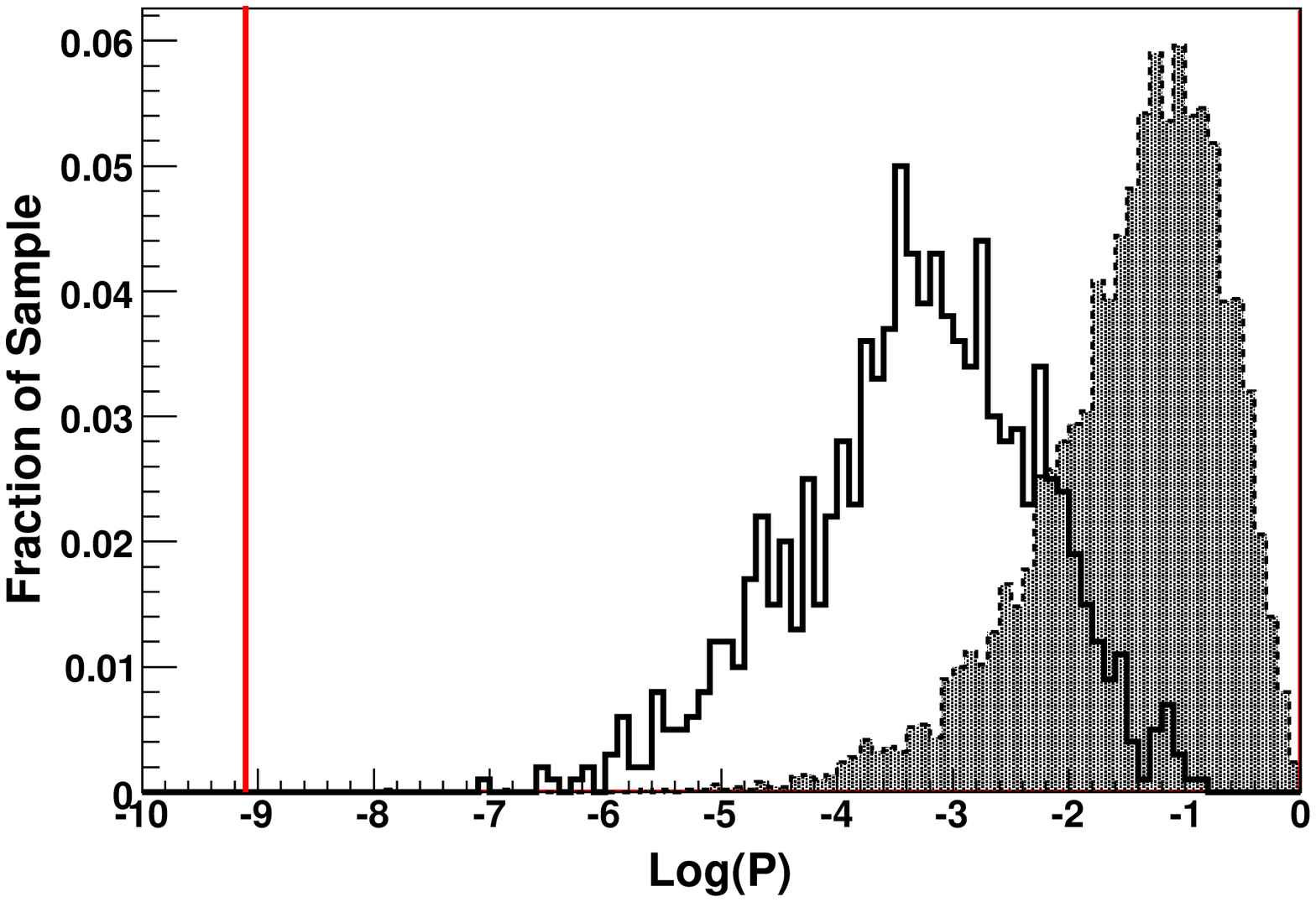,width=3in}
\epsfig{file=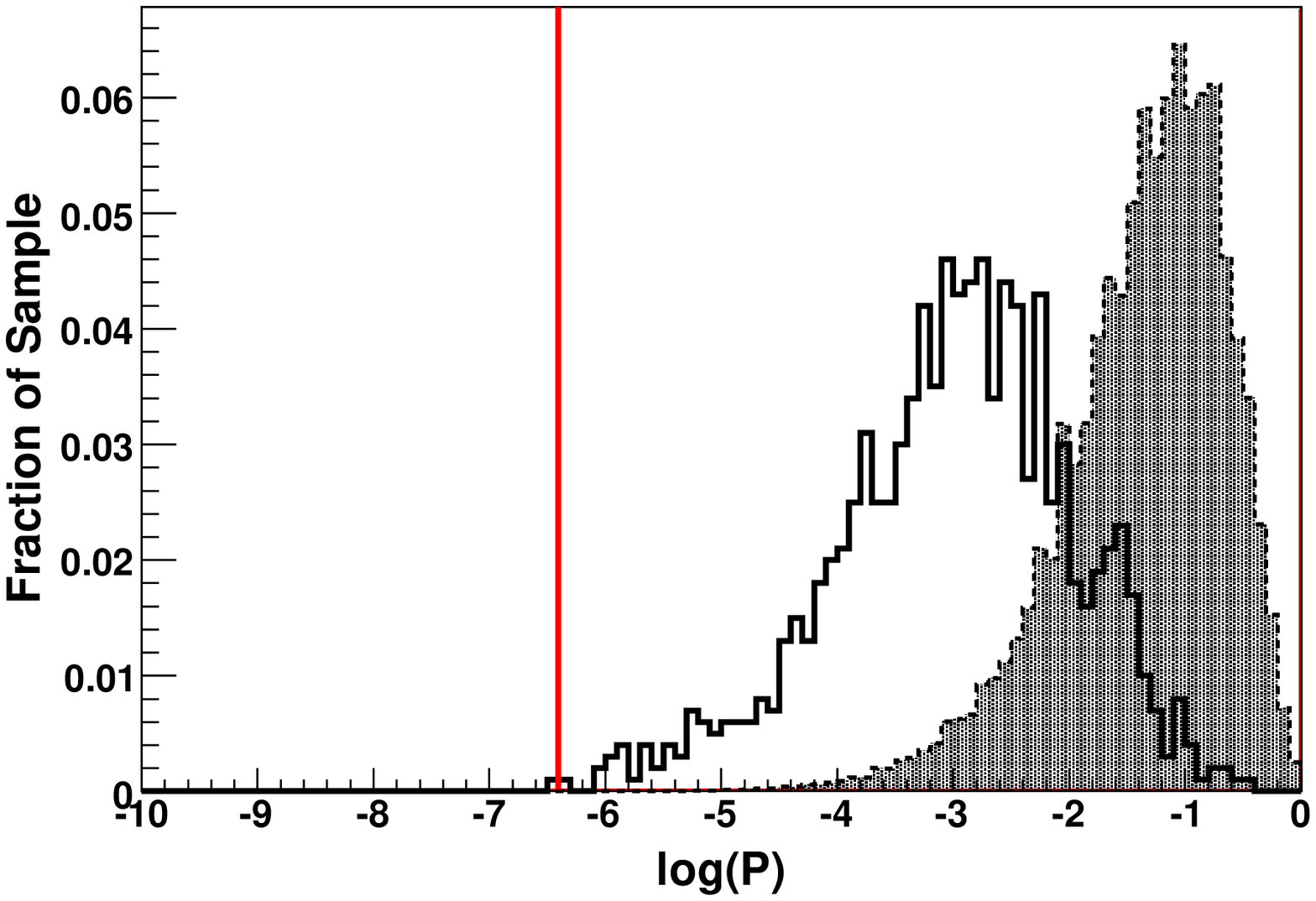,width=3in}
\end{center}
\caption{
Upper panel:  Distribution of minimum $P$ values for the 22 published Auger events with $|b| \geq 10^\circ$ in $\psi-z_{\rm max}$ scans assuming an isotropic source distribution ($10^6$ trials, grey histograms), and 1000 different realizations of 2MRS-based source catalogs (open histograms).  The minimum $P$ value using the VCV catalog is shown as a vertical line.  The lower panel is the same but for the cleaned VCV catalog, cVCV.}
\label{2MASS}.
\end{figure}

We next estimate the effect on the UHECR-VCV correlation, of restricting to galaxies in VCV which are confirmed AGNs, quasars or BLLacs. The primary effect of removing galaxies not so identified from the VCV catalog comes from correcting the classifications of VCV galaxies close to UHECRs; a secondary effect is that the cleaned source catalog is sparser, reducing $p$.  

To create a ``cleaned VCV" catalog (cVCV) we begin by removing the 14 not-optically-confirmed-AGNs with $z \leq 0.024$ within $6^\circ$ of the Auger UHECRs.  Then, we ``statistically clean" the remainder of the 628=694-66 VCV galaxies with $z \leq 0.024$ which are more than $6^\circ$ from a UHECR as follows.  We keep the 167 galaxies labeled S1.x and we remove all 66 galaxies labeled as H2.  That leaves 381 galaxies labeled S2, ?, or unlabeled, of which we select a fraction 33/42 at random, corresponding to the fraction of optically identified S2.x's in the same categories in the 66 VCV galaxies within $6^\circ$ which we examined individually.  This leaves us with 533 galaxies. After requiring $|b| \geq 10^{\circ}$, our final cVCV sample contains a total of 516 galaxies.  This procedure assumes that the statistics of the classification errors in the 66 galaxies within $6^\circ$ of the Auger UHECRs are representative of the ensemble.  While approximate, it gives quite an accurate indication of the correlation between optically-confirmed AGNs and UHECRs since it is exact for the galaxies close enough to produce a correlation and only the statistical properties of the remainder matter.  

Repeating the scan with this cleaned VCV catalog, we find that the most significant signal arises for $z_{\rm max} = 0.015$ and $\psi = 4.8^\circ$, for which 18 UHECRs are correlated, $p=0.2803$, and $P=2.4 \times 10^{-7}$.   The probability of getting a $P$ value this small or smaller if UHECRs were isotropic is $(2.9 \pm 0.5) \times 10^{-5}$.  The shaded histograms in Fig. \ref{2MASS} show the distributions of $P$ values over $10^6$ different isotropic samples, for the VCV (upper panel) and cVCV (lower panel) cases.  The values for the real VCV and cVCV correlation analyses are shown as vertical lines for purpose of comparison.   Thus the correlation between UHECRs and the cleaned VCV catalog is significantly stronger than for an isotropic source distribution, but less so than with the full VCV catalog.   

\section{Do VCV galaxies merely "trace" the sources of UHECRs?}
\label{2MRS}

In this section we test whether the correlation of the highest energy UHECRs with VCV can result from VCV galaxies being ``tracers" of the true sources.  AGNs are in first approximation a random subset of ordinary galaxies -- indeed, all galaxies may undergo AGN phases as a result of mergers with another galaxy which stimulates accretion.  An important question is therefore whether the 3.2$^\circ$ correlation scale might just reflect the galaxy clustering scale, inducing a similar correlation between UHECRs and any random sample of galaxies of the same size and redshift distribution as VCV.   
  
 \begin{figure}[t]
\begin{center}
\noindent
\epsfig{file=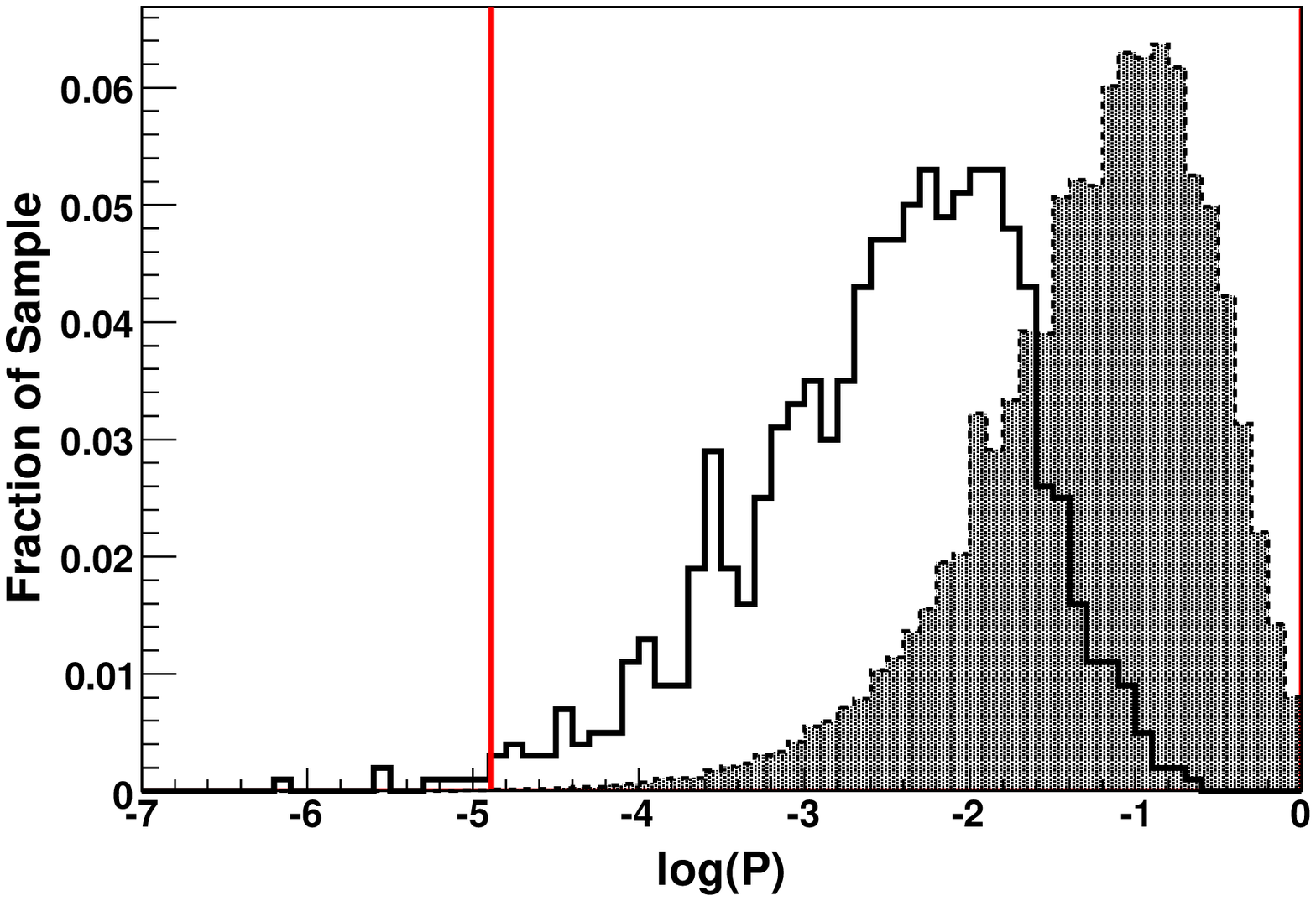,width=3in}
\epsfig{file=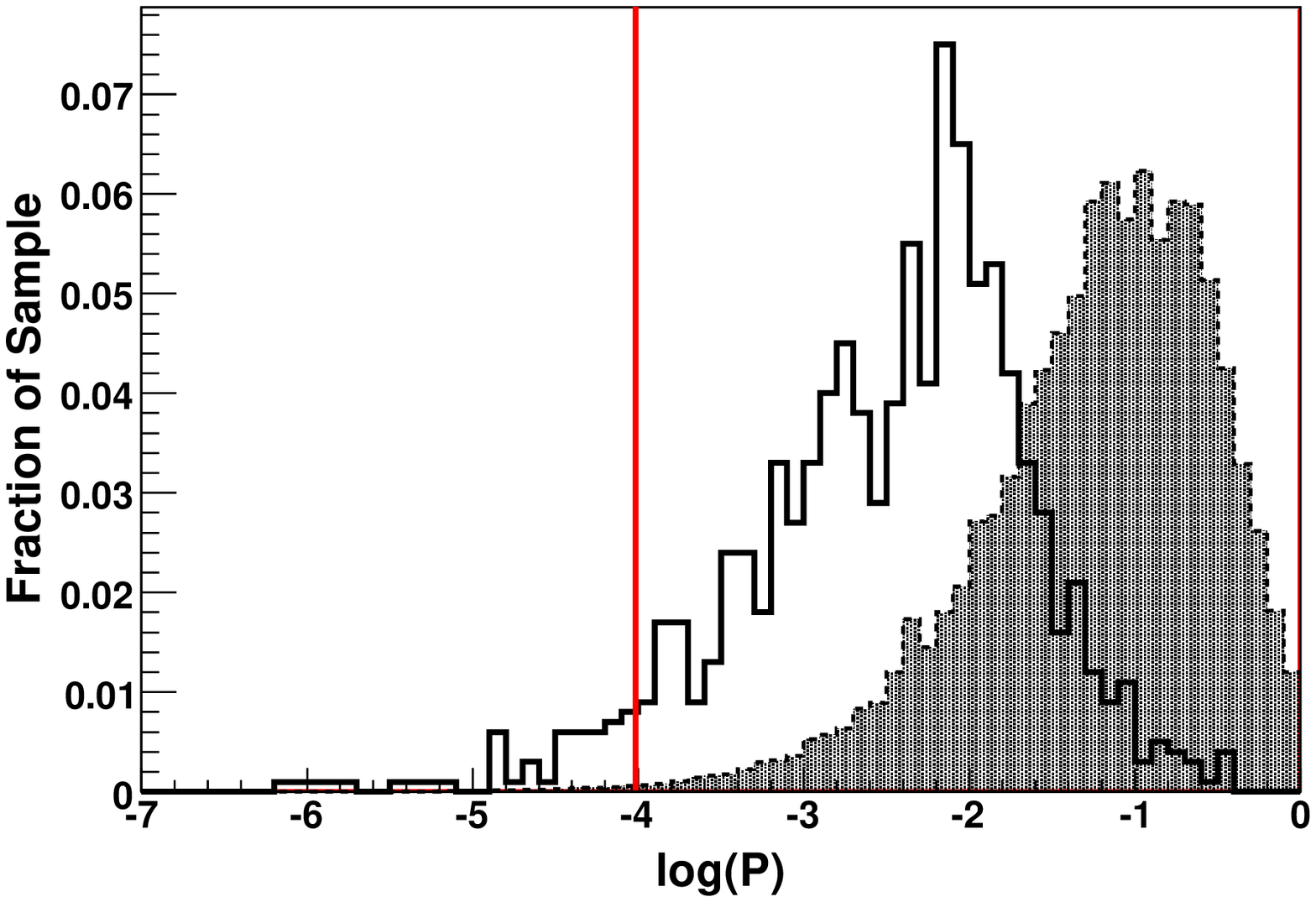,width=3in}
\end{center}
\caption{
As in Fig. \ref{2MASS}, but restricting to the 11 post-prescription UHECR events.  Only 0.7\% of the trials with 2MRS-based catalogs have as strong or stronger correlation than with VCV (upper panel) but this increases to $\approx 5$\% for cVCV (lower panel).}
\label{post}
\end{figure}

The analysis we perform to answer this question is straightforward: repeat the correlation scan analyses, but with 2MRS galaxies substituting for VCV or cVCV galaxies, and determine how often a correlation as strong as that observed is obtained.  We create 1000 different samples containing 674 (516) galaxies randomly chosen from 2MRS galaxies\footnote{We have removed the VCV galaxies from the 2MRS sample to avoid a correlation induced by the VCV content; this makes only an imperceptible difference in the numbers.}, with the same $z$ distribution as the VCV (cVCV) sample.  For each galaxy sample, we repeat the same correlation analysis as described above, for VCV (cVCV) galaxies, finding for each galaxy sample the value of $z_{max}$ and $\psi$ that minimizes the  probability measure $P$.  The open histograms in Fig. \ref{2MASS} show the resulting distributions of $P$ values over the 1000 different 2MRS samples, for the VCV (upper panel) and cVCV (lower panel) cases.  The values for the real VCV and cVCV correlation analyses are shown as vertical lines for purpose of comparison.   

Fig. \ref{2MASS} shows that -- not surprisingly -- UHECRs correlate more strongly with 2MRS galaxies than they do with the isotropic samples.  Nonetheless, the correlations between UHECRs and VCV or cVCV are significantly stronger than with 2MRS galaxies:  not one of our 1000 mock 2MRS samples has a $P$ value smaller than that for VCV, and only one has a lower value than the $P$ in the corresponding 1000 cVCV trials.  

It is known that galaxy clustering depends on the mass and luminosity of the sample.  Therefore, we repeated the analysis using the subsample with $M_K \leq -23.8$, which is volume-limited out to $z \leq 0.024$, instead of the 2MRS flux-limited sample.  The results do not change significantly.  The results of the mock catalog analyses are also essentially the same whether the mock catalogs are chosen to have the same number of galaxies as VCV (cVCV) in redshift bins of $\Delta z = 0.002$, or just the same total number with $z \leq 0.024$.   

Since the interest in the correlation of UHECRs with VCV galaxies and the choice of 57 EeV cutoff resulted from the analysis of the pre-perscription events (those taken prior to May 28, 2006), in order to assess the true unbiased significance of the correlation we must restrict to the post-prescription UHECR dataset (taken between May 28, 2006 and Aug. 31, 2007) and repeat the analysis described above with these 11 events only.  The results are shown in Fig. \ref{post}.  Scanning with the VCV galaxies,  the lowest value of $P$ is for $ z_{\rm max} = 0.014$ and $\psi_{\rm max} = 3.3^\circ$.  For these parameters, 9 UHECRs are correlated and $p = 0.1908$, giving $P = 1.3 \times 10^{-5}$.  Running the scan using $10^6$ mock isotropic catalogs and $10^3$ 2MRS realizations, the chance probability of finding as low or lower a value of $P$ from isotropic sources is $(4.6 \pm 0.2) \times 10^{-4}$ and from 2MRS-distributed sources is 0.7\%.   The same analysis for cVCV gives $ z_{\rm max} = 0.009$, $\psi_{\rm max} = 5.1^\circ$; 8 UHECRs are correlated, $p = 0.1781$, $P = 1.0 \times 10^{-4}$, and the chance probability is $(3.2 \pm 0.1) \times 10^{-3}$ and 5.3\%, for isotropic and 2MRS-subsampled sources.  The correlation is weaker for both VCV and cVCV with post-prescription UHECRs than with the full UHECR dataset, but this is normal due to the smaller number of events, and the fact that the 57EeV cutoff was chosen to maximize the correlation with pre-perscription events.  

The analysis presented here shows that the observed correlation between published UHECRs and VCV galaxies has only a very small chance of arising from galaxy clustering alone, implying that at least some of the 22 Auger UHECRs with $E > 57$ EeV and $|b| \geq 10^{\circ}$ are produced in galaxies in the VCV catalog with $z \leq 0.018$.   This result is not in contradiction with the report\cite{kashtiWaxman08} of finding no excess correlation between UHECRs and VCV galaxies compared to that arising from UHECR sources following the distribution of matter, because the method used in \cite{kashtiWaxman08} is intrinsically less sensitive than the present one and has insufficient power to discriminate.  

\section{ Conclusions}
We measured the completeness of the Veron-Cetty Veron AGN catalog and find that it varies strongly as a function of location on the sky and redshift.  Thus the VCV catalog cannot be used to determine the fraction of UHECRs that are produced by AGNs.  Furthermore, efforts to test or refute the UHECR-AGN correlation by comparing the {\em global} distribution of UHECRs to that of VCV galaxies are not valid.  A uniform and well-characterized AGN catalog will be essential for that purpose.

We followed up on the study of Zaw et al\cite{zfg09}, which found that only 2/3 of the VCV galaxies within 3.2$^\circ$ of a UHECR are optically identifiable AGNs, by investigating whether there is a significant correlation when the possibly-not-AGNs are removed from the source catalog.  We find that the correlation is reduced but remains significant.  It will be interesting to find out whether some of the optically-not-identified-as-AGN galaxies in VCV do show AGN activity in radio or X-ray, or whether another type of source is needed in addition to AGNs.

Finally and most importantly, we have introduced a simple but effective way to determine whether the correlation observed between the highest energy cosmic rays and galaxies in the Veron-Cetty Veron AGN catalog may simply result from the fact that AGNs are galaxies and galaxies are clustered.  If this were the source of the UHECR-VCV correlation, then repeatedly subsampling a complete catalog of galaxies in the same redshift range to a similar density would regularly produce a comparable degree of correlation.  Instead, we find that a correlation as strong as what is observed between the {\em post-prescription} (May 28, 2006-Aug. 31, 2007) Auger UHECRs with $|b|>10^\circ$ and energy above 57 EeV, and galaxies in the VCV catalog, occurs in fewer than 1\% of randomly drawn subsamples of either volume-limited or flux-limited 2MRS galaxy catalogs.  As expected, such subsamples of galaxies do typically have a stronger correlation with UHECRs than found for an isotropic source distribution, however the increase in the correlation is too small to explain the extent of the correlation observed with VCV.   Thus we conclude that either the degree of correlation between highest energy Auger UHECRs and VCV galaxies in the post-prescription events is a statistical fluke or some of these cosmic rays are in fact produced in VCV galaxies.

We acknowledge the essential role of the Pierre Auger Collaboration in this work, which is based on data obtained and published by the  Pierre Auger Observatory.  Furthermore, GRF and IZ acknowledge their membership in the Pierre Auger Collaboration and thank their colleagues for their participation in and contribution to this research.  We especially thank J. Huchra and members of the 2MRS team, for pre-release use of the 2MASS Redshift Survey K=11.25 catalog.  Some of the results in this paper have been derived using the HEALPix\cite{healpix} package.  This research has been supported in part by NSF-PHY-0701451.

%\bibliographystyle{unsrt}
%\bibliography{mnem,CR,AGN,uhecr}

\end{document}